\documentclass[aip,amsmath,amssymb,reprint,]{revtex4-1}

\usepackage{graphicx}
\usepackage{dcolumn}
\usepackage{bm}
\usepackage[utf8]{inputenc}
\usepackage[T1]{fontenc}
\usepackage{mathptmx}

\draft 

\begin{document}

\title{Optomechanical response with nanometer resolution in the self-mixing signal of a terahertz quantum cascade laser} 

\author{Andrea Ottomaniello}
\affiliation{Dipartimento di Fisica "E. Fermi", Università degli studi di Pisa, 561217 Pisa, Italy}
\affiliation{NEST, CNR -Istituto Nanoscienze and Scuola Normale Superiore, Piazza San Silvestro 12, 56127 Pisa, Italy}

\author{James Keeley} 
\affiliation{School of Electronic and Electrical Engineering, University of Leeds, Leeds LS29JT, United Kingdom}

\author{Pierluigi Rubino}
\affiliation{School of Electronic and Electrical Engineering, University of Leeds, Leeds LS29JT, United Kingdom}

\author{Lianhe Li} 
\affiliation{School of Electronic and Electrical Engineering, University of Leeds, Leeds LS29JT, United Kingdom}

\author{Marco Cecchini}
\affiliation{NEST, CNR -Istituto Nanoscienze and Scuola Normale Superiore, Piazza San Silvestro 12, 56127 Pisa, Italy}

\author{Edmund H. Linfield}
\affiliation{School of Electronic and Electrical Engineering, University of Leeds, Leeds LS29JT, United Kingdom}

\author{A. Giles Davies}
\affiliation{School of Electronic and Electrical Engineering, University of Leeds, Leeds LS29JT, United Kingdom}

\author{Paul Dean}
\affiliation{School of Electronic and Electrical Engineering, University of Leeds, Leeds LS29JT, United Kingdom}

\author{Alessandro Pitanti}
\affiliation{NEST, CNR -Istituto Nanoscienze and Scuola Normale Superiore, Piazza San Silvestro 12, 56127 Pisa, Italy}

\author{Alessandro Tredicucci}
\affiliation{Dipartimento di Fisica "E. Fermi", Università degli studi di Pisa, 561217 Pisa, Italy}
\affiliation{NEST, CNR -Istituto Nanoscienze and Scuola Normale Superiore, Piazza San Silvestro 12, 56127 Pisa, Italy}

\date{\today}

\begin{abstract}
The effectiveness of self-mixing interferometry has been demonstrated across the electromagnetic spectrum, from visible to microwave frequencies, in a plethora of sensing applications, ranging from distance measurement to material analysis, microscopy and coherent imaging. Owing to their intrinsic stability to optical feedback, quantum cascade lasers (QCLs) represent a source that offers unique and versatile characteristics to further improve the self-mixing functionality at mid-infrared and terahertz (THz) frequencies. Here, we show the feasibility of detecting with nanometer precision deeply subwalength ($<\lambda$/6000) mechanical vibrations of a suspended Si$_3$N$_4$-membrane used as the external element of a THz QCL  feedback interferometric apparatus. Besides representing a platform for the characterization of small displacements, our self-mixing configuration can be exploited for the realization of optomechanical systems, where several laser sources can be linked together through a common mechanical microresonator actuated by radiation pressure.  
\end{abstract}


\maketitle 

The self-mixing (SM) effect describes the mixing of the intracavity electromagnetic field of a laser with its emitted radiation partially reinjected into the laser cavity \cite{kane2005unlocking}. Although such optical feedback (OF) can be detrimental to laser operation \cite{kleinman1962discrimination}, 
the SM effect can also be exploited for metrological applications  \cite{king1963metrology}, through a technique known as laser-feedback interferometry (LFI) \cite{taimre2015laser}. This homodyne technique, in which the laser acts simultaneously as source, mixer and shot-noise limited detector, allows retrieval of information about the external cavity, comprising the target and the external medium, by monitoring the response of the laser to OF. Thanks to the universal and simple character of the SM phenomenon, its functionality has been demonstrated from the visible to the microwave range using laser systems as diverse as gas lasers \cite{robert1968apparatus}, solid-state lasers \cite{nerin1997self} and semiconductor lasers \cite{seko1975self}. This has led to the development of a large number of sensing applications \cite{bosch2001optical, giuliani2002laser} ranging from displacement measurement \cite{donati1995laser} to material analysis \cite{rakic2013swept}, laser emission spectrometry \cite{keeley2017measurement} and coherent imaging \cite{dean2013coherent, wienold2016real}.  
Among all kinds of semiconductor lasers, quantum cascade lasers (QCLs) allow further simplification of the SM scheme thanks to their intrinsic voltage sensitivity to OF. This allows the SM modulation to be measured directly, and with high sensitivity \cite{keeley2019}, via the voltage variation across the active region with no need for an external photodector \cite{dean2011terahertz}. Moreover, QCLs exhibit peculiar ultra-stability to OF\cite{mezzapesa2013intrinsic}, due to their small linewidth enhancement factor \cite{green2008linewidth} ($0<|\alpha|<1$) and long photon to carrier lifetime ratio. These unique and versatile characteristics offer an opportunity for the development of high performance LFI schemes operating at mid-infrared and terahertz (THz) frequencies, for coherent imaging \cite{dean2014terahertz} and displacement sensing \cite{leng2011demonstration, mezzapesa2014qcl}. In fact, despite the relatively long wavelength, nanometer target displacements as low as $\lambda/100$ have been measured by adding a fast moving etalon in the external cavity \cite{mezzapesa2015nanoscale}.

In this letter, we report a significant improvement of nanometer displacement sensing by showing the detection of deeply subwavelength vibrations ($< \lambda/6000$) of a suspended mechanical resonator, by employing a 3.34 THz QCL operating in continuous-wave (CW). The experimental results have been validated by solving the Lang-Kobayashi (LK) model \cite{lang1980external} in the steady-state regime after calibration of the absolute membrane position. Furthermore, the measured oscillation amplitudes are shown to be in agreement with independent measurements performed using a laser Doppler vibrometer (LVD) operating in the visible region.

\begin{figure}[ht!]
\centering\includegraphics[width=4.5 cm]{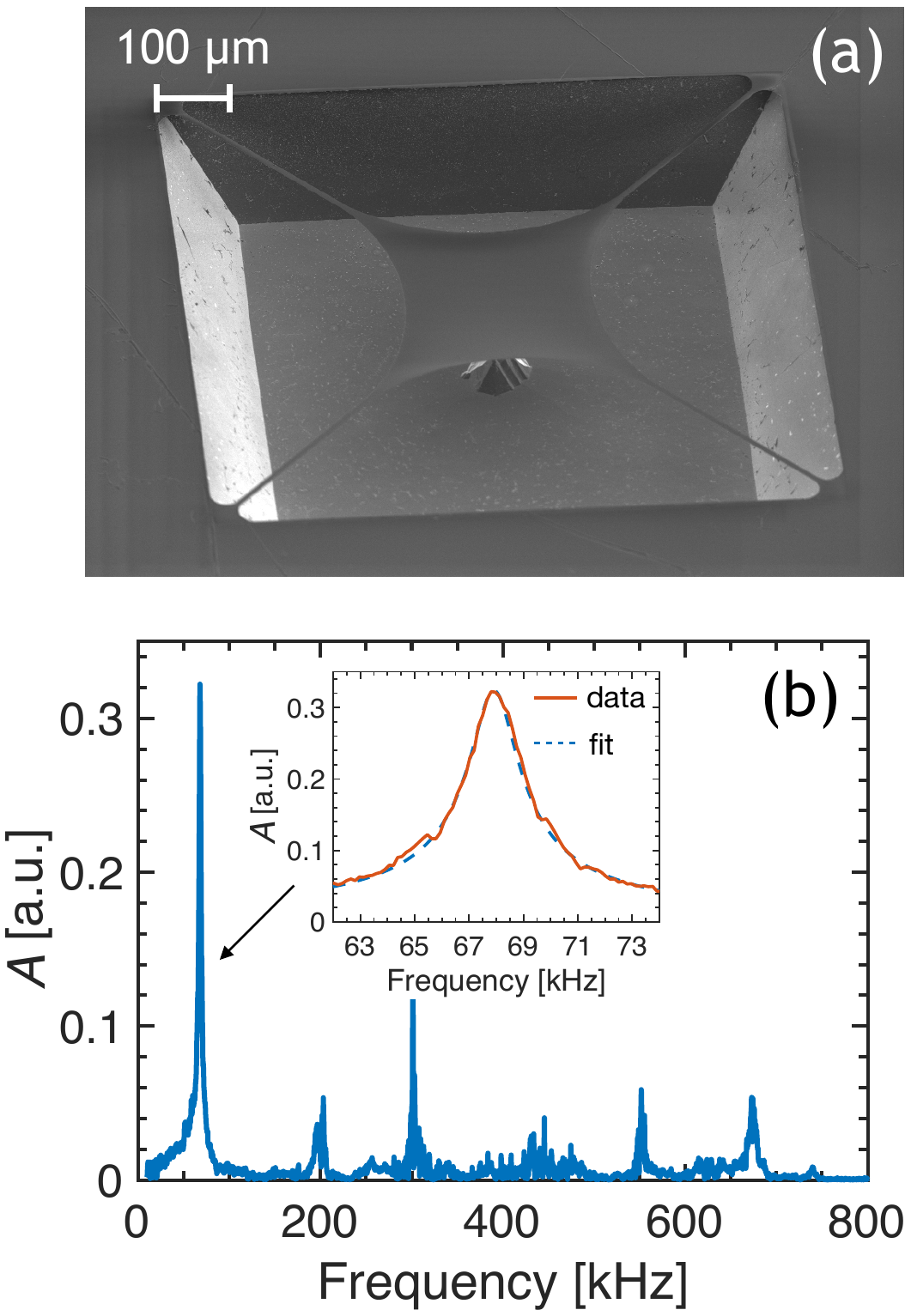}
\caption{(a) SEM image of the Si$_3$N$_4$ membrane. (b) Membrane mechanical spectrum measured at atmospheric pressure at the membrane center using the LDV. Inset: zoom of the fundamental resonance and corresponding fit (dashed line) using Eq. 2.} \label{fig_1}
\end{figure}

We chose as an external element a Si$_3$N$_4$ trampoline membrane, which is an optimum candidate for interferometric readout and optomechanical applications \cite{reinhardt2016ultralow, norte2016mechanical}. Such membranes have already been used to observe optomechanical features in an infrared laser diode SM configuration \cite{baldacci2016thermal}. \\ A SEM image of the fabricated sample is shown in \figurename~\ref{fig_1}(a). 

The structure is composed of a 300-nm-thick, 300-$\mu$m-wide suspended central pad held at each corner by 10-$\mu$m-wide tethers anchored at the vertices of a 1-mm-side squared window. Details of the sample fabrication can be found elsewhere \cite{baldacci2016thermal}. The membrane pad dimensions were chosen in order to increase the level of OF for SM measurements by creating a reflective surface of size comparable to the focused THz beam spot size ($\sim250$ $\mu$m). To further increase the reflectivity, a 10-nm-thick gold layer was thermally evaporated over the entire membrane surface to produce an almost totally reflective target. The sample was directly glued on the top surface of a piezoelectric ceramic actuator with the fundamental out-of-plane mechanical resonance at $\sim100$ kHz. By exciting the piezo-actuator, the membrane motion was driven and its mechanical response could be investigated. The system composed of the piezo-actuator and membrane sample was mounted inside a vacuum chamber allowing control of the environmental pressure. An initial membrane characterization was performed using a commercial high spatial-resolution LDV (Polytech UHF 120). By applying a flat spectrum voltage excitation to the piezo-actuator, the resonant mechanical modes of the membrane could be observed.
 
\begin{figure}[ht!]
\centering\includegraphics[width=6cm]{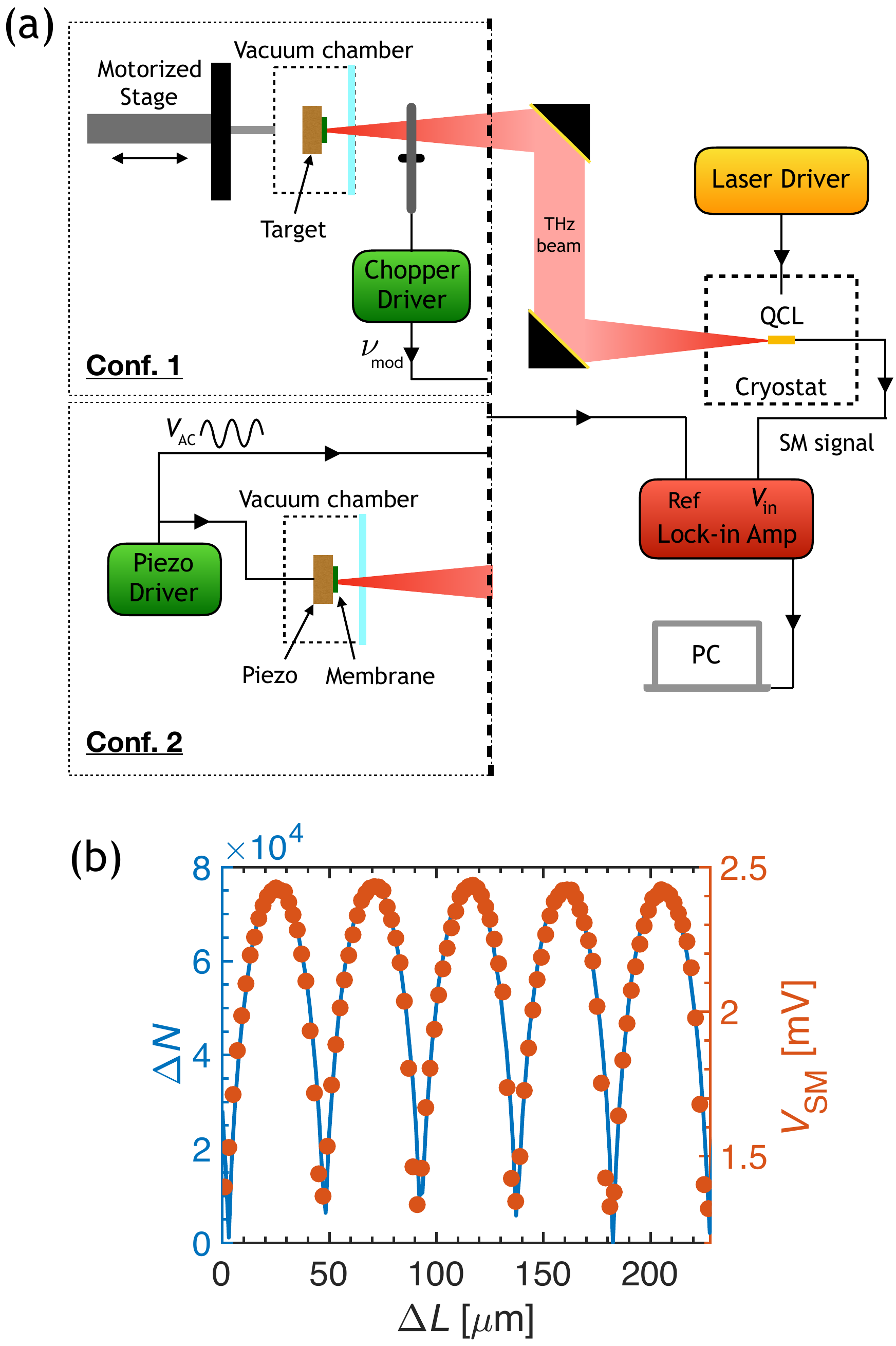}
\caption{(a) Sketch of the two configurations of the SM apparatus. 
(b) Calculated $\Delta N$ (blue curve), and measured $V_{\text{SM}}$ (red points), as a function of $\Delta L$ using configuraton 1.} \label{fig_2}
\end{figure}

From the spectrum shown in \figurename~\ref{fig_1}(b), obtained at atmospheric pressure, the fundamental mechanical mode is observed to be well separated from the higher order modes. Moreover, with the membrane oscillating orthogonally to the chip-plane, this particular mode ensures that the full THz wavefront reflected from the membrane experiences the same external optical path, thus contributing with the same phase to the SM effect. This makes the fundamental vibrational mode particularly suitable for coherent measurements of membrane vibrations by LFI \cite{baldacci2016thermal}and, as such, subsequent measurements here focus on this mode. A complete LDV characterization of the membrane motion as a function of pressure was performed, and is reported for convenience in the Supplementary Material. The data was analysed by modelling the membrane as a one-dimensional driven harmonic oscillator. Defining $A(t)$ and $a(t)$ as the positions of the membrane trampoline and of the tether clamps (moving with the piezo driver) with respect to their common rest positions $A=a=0$, respectively, the membrane equation of motion can be written as:
\begin{equation}
\ddot A(t) + \gamma \dot A(t) + \omega_{\text{m}}^2 [A(t)- a(t)] = 0  \label{eq.1}
\end{equation}
where $\omega_{\text{m}}$ is the pressure-dependent membrane resonance angular frequency and $\gamma$ is the membrane damping coefficient. The oscillation amplitude can be obtained from the imaginary part of the solution, $A(\omega)$, in the frequency domain:
\begin{equation}
\text{Im} [A(\omega)] = \frac{\gamma \omega}{(\omega_{\text{m}}^2-\omega^2)^2+ (\gamma \omega)^2}  \omega_{\text{m}}^2 a_0\label{eq.2}
\end{equation}
where $a_0$ is the maximum driving oscillation amplitude. \\ By fitting the displacement spectra using Eq. 2, we obtain for each investigated pressure the values of the resonance frequency $\omega_\text{m}/2 \pi$, the quality factor $Q=\omega_{\text{m}}/\gamma$ and the oscillation amplitude at the resonance frequency $A^{\text{max}}$. As the environmental pressure decreases these three quantities monotonically increase up to a saturation level. The resonance frequency and quality factor change from $\omega_{\text{m}}/2\pi\sim67.9$ kHz and $Q\sim40\pm2$, respectively, at atmospheric pressure (see inset of \figurename~\ref{fig_1}(b)) to $\sim73.4$ kHz and $\sim200\pm10$ at pressure $\sim7.3\times10^{-3}$ mbar. 

A schematic diagram of the experimental apparatus for the THz LFI measurements is presented in \figurename~\ref{fig_2}(a). We used a THz QCL consisting of a $14$-$\mu$m-thick GaAs-AlGaAs active region with growth details reported in the literature \cite{wienold2009low}. The device was fabricated as a single-metal ridge with longitudinal and transversal dimensions of 1.8 mm and 150 $\mu$m, respectively. The laser was driven in CW mode with a current equal to 510 mA and was maintained at a temperature of $25 \text{K}$ using a continuous-flow cryostat. Under these conditions the QCL provided stable single mode emission at 3.34 THz and a large SM response \cite{wienold2009low, keeley2019}. The THz beam passing through the polythene window of the cryostat was collimated and focused (using two $f/2$ off-axis parabolic reflectors with diameter 50.8 mm) onto the trampoline membrane which constituted the mirror of the external cavity providing OF to the laser. The vacuum chamber containing the membrane was fixed to a motorized stage allowing the external optical cavity length $L$ between the target and the QCL emission facet to be micrometrically changed. The LFI apparatus was used in two configurations. In the first configuration (conf. 1 in \figurename~\ref{fig_2}(a)), the membrane motion was not excited and the membrane was translated towards the laser facet in 2 $\mu$m-long steps parallel to the beam propagation direction. The QCL SM voltage $V_{\text{SM}}$ was measured by a lock-in amplifier synchronized to the optical modulation frequency ($\nu_{\text{mod}} = 212$ Hz) imposed to the THz beam by a mechanical chopper placed in front of the vacuum chamber. In the second configuration (conf. 2 in \figurename~\ref{fig_2}(a)) the membrane equilibrium position was instead kept fixed, but the membrane vibrated at the frequency of the piezo driving excitation, in turn used as the reference of the lock-in amplifier. 

To model the SM voltage signal as a function of the target position we used the steady-state solutions of the Lang-Kobayashi equations describing the laser under OF \cite{lang1980external}. This was done by evaluating the laser emission frequency under OF, $\omega_{\text{F}}$, by numerically solving the so-called excess phase equation: 
\begin{equation}
\omega_0-\omega_{\text{F}}=\frac{k}{\tau_{\text{ext}}} \sqrt{1 + \alpha^2} \sin (\omega_{\text{F}} \tau_{\text{ext}} + \arctan \alpha) \label{eq. 3}
\end{equation}
where $\omega_0$ is the laser emission frequency without feedback, $\tau_{\text{ext}}=2L/c$ is the round-trip time in the external cavity, $\alpha$ is Henry's linewidth enhancement factor and $k$ is the OF coupling rate. Using the conventional three mirrors model in the weak feedback regime, where only one reflection from the target is considered, $k$ can be written as $k=\epsilon \sqrt{\frac{R_{\text{ext}}}{R}} (1- R)$, 
in which $\epsilon$, $R$ and $R_{\text{ext}}$ are the coupling-efficiency factor and the reflectivities of the target and laser emission facet, respectively. 

\begin{figure}[ht!]
\centering\includegraphics[width=5.5cm]{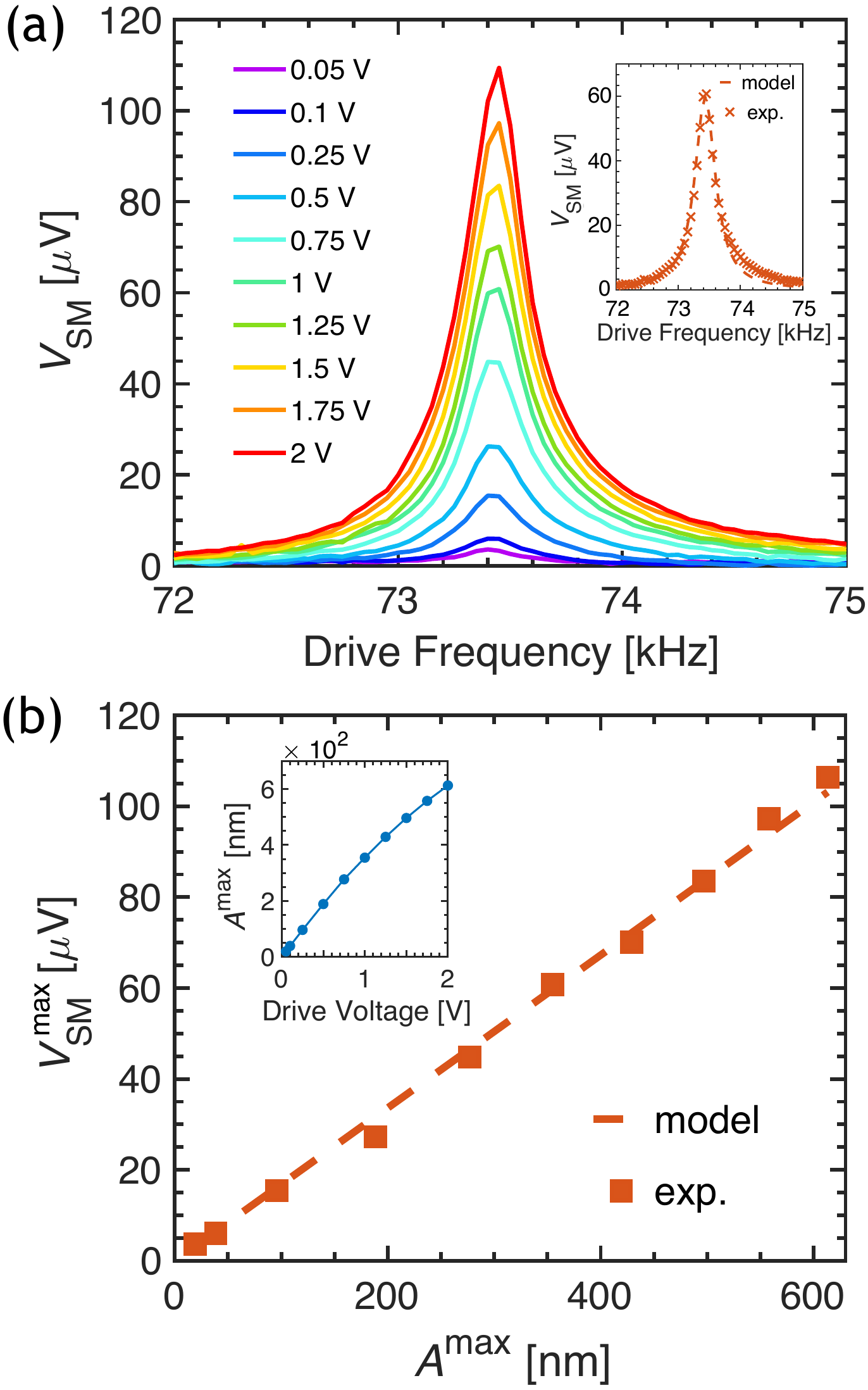}
\caption{(a) Measured $V_{\text{SM}}$ as a function of the drive frequency for different applied voltages at $P=0.5$ mbar. Inset: computed (red dashed line) and experimental (red crosses) $V_{\text{SM}}$ spectrum with 1$V_{\text{RMS}}$ applied to the piezo actuator. (b) Measured (red squares) and calculated (dashed line) $V^{\text{max}}_{\text{SM}}$ as a function of $A^{\text{max}}$. Inset: LDV-measured $A^{\text{max}}$ corresponding to the experimental points in the main graph for the same biases and pressure reported in (a). }\label{fig_3}
\end{figure}

The parameter $\omega_{\text{F}}$ obtained from Eq. \ref{eq. 3} determines the laser carrier number variation $\Delta N$ from the threshold carrier number with and without feedback via the following equation:
\begin{equation}
\Delta N = - \frac{2 k}{G \tau_{\text{c}}} \cos (\omega_{\text{F}} \tau_{\text{ext}}) \label{eq. 4}
\end{equation}
where $G$ is the active region modal gain factor and $\tau_{\text{c}}$ the laser cavity round-trip time. 
Starting from the initial external cavity optical length, $L_0=47$ cm, $\Delta N$ was calculated for each value of $L$ using the best fitting parameters reported in Supplementary Material. The values of $\Delta N$ are plotted (blue curve) as a function of the membrane displacement $\Delta L=L_0-L$ in \figurename~\ref{fig_2}(b) together with the experimental values $V_{\text{SM}}$ (red points) obtained with configuration 1. The theoretical values are in good agreement with the experiment reproducing both the same shape and typical periodicity ($\lambda_{F}$/2) of the SM signal. From comparison of these curves we obtained the proportionality coefficient $\beta$ between the calculated $\Delta N$ and the measured $V_{\text{SM}}$, found to be $3.1\times10^{-8}$ mV. 
It should be noted that the calibration factor $\beta$ depends on the particular value of $G$, which was fixed as $1.42\times10^{4}$ s$^{-1}$ according to common values in THz QCL literature \cite{agnew2016model, hamadou2008modelling}. Nevertheless, our calibration procedure allows us to correlate experiments and simulations independently of the value assigned to $G$.

SM measurements of the membrane oscillation amplitude were then performed using configuration 2. $V_{\text{SM}}$ measured at an environmental pressure of $P=0.5$ mbar is reported as a function of the drive frequency and for several RMS drive voltages in \figurename~\ref{fig_3}(a). 
Using a Lorentzian spectral line-shape as the fitting function for the data (valid in the experimentally verified high-$Q$ limit), the vibration spectra were observed to have the same resonance frequency $\omega_{\text{m}}/2\pi\sim73.43\pm0.01\ $kHz and $Q\sim198\pm2$, independent of the bias applied to the piezo-actuator. These quantities agree with those obtained previously with the LDV.  
The $V_{\text{SM}}$ signal measured in response to the membrane vibrations can be modelled with Eqs. \ref{eq. 3}-\ref{eq. 4} for the steady-state solution of the LK model. In fact, with the limit $2\pi/\tau_{\text{ext}}>>\omega_{\text{m}}$ satisfied, the membrane position at each instant in time can be considered fixed, contributing as a static external optical path for the LK equations. The time-dependent variation of carrier number $\Delta N(t)$ corresponding to a certain membrane oscillation amplitude can thus be numerically obtained by inserting into Eqs. \ref{eq. 3}-\ref{eq. 4} the following expression for $L$:
\begin{equation}
L=L_0+A(\omega) \cos (\omega_{\text{m}} t) 
\end{equation}
where $A(\omega)$ is the drive frequency-dependent membrane oscillation amplitude (Eq. \ref{eq.2}) evaluated using the previously fitted values of $\gamma$ and $\omega_{\text{m}}$. The total carrier number variation corresponding to the membrane vibration at a given drive frequency, $\Delta N(\omega)$, can then be calculated as the peak-to-peak amplitude of the time-oscillating $\Delta N(t)$. $V_{\text{SM}}$ can then be obtained just multiplying $\Delta N(\omega)$ by the previously retrieved $\beta$-factor. In the inset of \figurename~\ref{fig_3}(a), the values of $V_{\text{SM}}$ resulting from this model are shown to reproduce well the experimental $V_{\text{SM}}$ measured by applying a drive bias of 1 $V_{\text{RMS}}$ spanning a single frequency at a time in the range 72$-$75 kHz. For that drive voltage both model and experiment result in a maximum SM signal of $\sim61\pm1$ $\mu$V obtained for a $A=355\pm5$ nm. Only a small deviation between data and model was found on the high-frequency side of the resonance, which we ascribe to a small degree of mechanical anharmonicity of the tethers' motion not included in the model. 
Both the theoretical and experimental maximum SM voltages, $V^{\text{max}}_{\text{SM}}$, are shown in \figurename~\ref{fig_3}(b) as a function of the membrane maximum oscillation amplitude, $A^{\text{max}}$. For the experimental points, the reported values of $A^{\text{max}}$ are those obtained from the LDV measurements at the same piezo-actuator drive voltage as the SM measurements, and they are shown in the inset as a function of the applied bias. From the main graph it can be observed that the model matches the experimental data and confirms a linear relation between $V^{\text{max}}_{\text{SM}}$ and $A^{\text{max}}$ with a slope coefficient of $0.17\pm0.01$ $\mu$V/nm. 
This agreement implies that we are able to detect membrane vibrations with appreciable precision ($\le$ 10 nm, limited by the voltage noise of our set-up) down to deeply subwavelength oscillation amplitudes of only a few nanometers. Specifically, the amplitude corresponding to a SM signal $\ge3$ dB above the voltage noise is $\sim15\pm5$ nm.

\begin{figure}[ht!]
\centering\includegraphics[width=\linewidth]{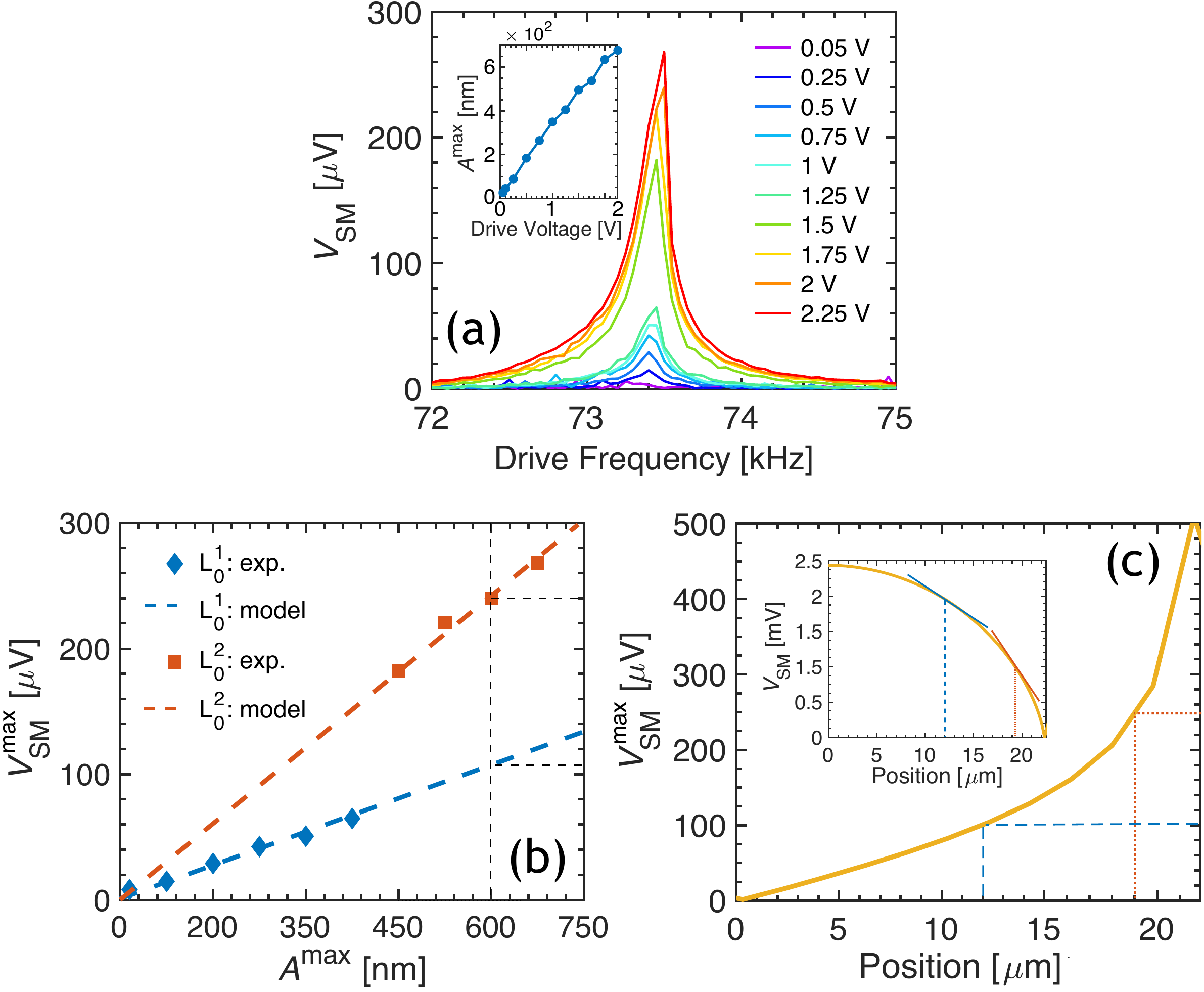}
\caption{(a) Measured $V_{\text{SM}}$ as a function of the drive frequency for different piezo applied voltages at $P=4.3\times10^{-3}$ mbar. Inset: LDV measured $A^{\text{max}}$ for same piezo drive voltages. (b) Blue diamonds and red squares: experimental $V^{\text{max}}_{\text{SM}}$ obtained from data in (a), as a function of $A^{\text{max}}$ obtained for drive voltages until 1.25 $V_{\text{RMS}}$ and for the range from 1.5 to 2.25 $V_{\text{RMS}}$, respectively. Blue and red dashed lines: calculated $V^{\text{max}}_{\text{SM}}$ for two different membrane positions, labeled $L^1_0$ and $L^2_0$. The vertical dashed line refers to $A^{\text{max}}$ considered in the adjacent graph. (c) Calculated $V^{\text{max}}_{\text{SM}}$ for $A=600$ nm for various equilibrium membrane positions. Blue dashed and red dotted lines indicate the membrane position and $V^{\text{max}}_{\text{SM}}$ corresponding to $L_0^1$ and $L_0^2$, respectively. Inset: calculated $\it{static}$ $V_{\text{SM}}$ in the same range. Vertical dashed and dotted lines represent the same positions. Blue and red solid lines highlight the corresponding different slopes of $V_{\text{SM}}$.} \label{fig_4}
\end{figure}

The same SM measurement was also performed at a lower pressure ($P\sim4.3\times10^{-3}$ mbar). The membrane vibration spectra are shown in \figurename~\ref{fig_4}(a), while the inset reports  the LDV measurements of the maximum membrane displacement as a function of the drive voltage. 
Although the maximum amplitudes of the membrane vibrations at $P=4.3\times10^{-3}$ mbar do not significantly vary from those at $P=0.5$ mbar, the values of $V^{\text{max}}_{\text{SM}}$ at lower pressure and for drive voltages  >1.5 $V_{\text{RMS}}$ almost double with respect to those obtained at higher pressure. 
This effect is shown in \figurename~\ref{fig_4}(b), in which the measured $V^{\text{max}}_{\text{SM}}$ are reported versus the experimental $A^{\text{max}}$ as blue diamonds and red squares for drive voltages lower and higher than 1.5 $V_{\text{RMS}}$, respectively. From the model it is possible to demonstrate that the two sets of experimental points arise from different external optical lengths, named $L_0^1$ and $L_0^2$. For drive voltages $\le1.5$ $V_{\text{RMS}}$, data are matched by a linear trend with the same slope coefficient as in \figurename~\ref{fig_3}(b), which results from the model using the membrane distance $L_0=L_0^1$. For voltages $> 1.5$ $V_{\text{RMS}}$, the slope increases to $0.39\pm0.01$ $\mu$V/nm corresponding to an external cavity optical length $L_0=L_0^2=L_0^1+7 \ \mu$m. This difference in slope efficiency (or equivalently the SM voltage sensitivity to displacement) can be explained by observing \figurename~\ref{fig_4}(c) in which the calculated $V_{\text{SM}}^{\text{max}}$ (corresponding to $A^{\text{max}}=600\pm5$ nm for a drive bias of 2 $V_{\text{RMS}}$) is plotted (solid curve) against the membrane position. As shown in the inset, the spatial range reported here is a quarter of the SM signal period, i.e. from the top of a fringe, at which the maximum $\textit{static}$ SM signal is measured (Position=0), to the bottom of the same fringe. It can be observed that the SM voltage sensitivity monotonically increases moving away from the position of maximum $\it{static}$ SM voltage. The lower and higher sensitivities evident in \figurename~\ref{fig_4}(c) correspond to a distance from the maximum $\it{static}$ SM voltage of $\sim12$ $\mu$m and $\sim19$ $\mu$m, respectively, which in turn agree with the corresponding slopes of the $\it{static}$ SM signal at these positions, highlighted in the inset. In fact, since we were considering displacements $A^{\text{max}}\ll \lambda/2$, the SM voltage sensitivity is expected to be proportional to the derivative of the $\it{static}$ SM signal. \figurename~\ref{fig_4}(c) thus reveals that in principle a sensitivity of $\sim0.83\pm0.02$ $\mu$V/nm can be achieved in our system by fixing $L=L_0\pm22$ $\mu$m such that the operating point is close to the minimum of the $\it{static}$ SM voltage (corresponding to the maximum slope).

In conclusion, we have experimentally demonstrated the detection of deeply subwavelength vibrations ($<\lambda/6000$) by LFI in the THz frequency range and using as an external element a suspended Si$_3$N$_4$ membrane. The SM voltage signal arising across the THz QCL active region in response to nanometer oscillations of the membrane was reproduced and uniquely determined by numerically solving the LK model in the steady state regime. The membrane oscillation amplitudes determined from the SM measurements were also found to be in agreement with those measured using a standard laser-Doppler vibrometric technique.
The described LFI apparatus can thus constitute a platform for the characterization of the mechanical response of a variety of systems with a precision comparable
to that achieved at
near-infrared or visible frequencies. Moreover, the proposed system can be employed and implemented for optomechanical applications where the suspended membrane is driven by the radiation pressure; the realization of an optomechanical system constituted by two different lasers coupled through the mechanical motion driven by radiation pressure represents a promising perspective.

\bibliography{biblio}

\end{document}